\documentstyle[emulateapj,epsf,amstex,astron_mlb]{article}
\lefthead{Balogh et al.}
\righthead{}

\begin{document}
\newcommand{\oii}{[OII]$\lambda$3727}
\newcommand{\ow}{$W_{\circ}($OII$)$}
\newcommand{\hd}{$W_{\circ}(H\delta)$}
\newcommand{\ebv}{$E_{\footnotesize \bv}$}
\newcommand{\br}{{\em B-R}}
\newcommand{\minit}{\dot{M}_{\rm init}}
\def\items{\hangindent=0.5truecm \hangafter=1 \noindent}
\submitted{ApJ, resubmitted Mar 31, 2000}
\title{The Origin of Star Formation Gradients in Rich Galaxy Clusters}

\author{Michael L. Balogh\altaffilmark{1} and Julio F. Navarro\altaffilmark{2}}
\affil{Dept of Physics and Astronomy, University of Victoria\\
P.O. Box 3055, Victoria, B.C. V8W 3P6, Canada}
\author{Simon L. Morris}
\affil{Dominion Astrophysical Observatory\\Herzberg Institute of Astrophysics,
National Research Council of 
Canada\\ 5071 W. Saanich Rd., Victoria, B.C. V8X 4M6, Canada} 

\altaffiltext{1}{Present address: Department of Physics, University of
Durham, South Road, Durham UK DH1 3LE\\email: M.L.Balogh@@durham.ac.uk}
\altaffiltext{2}{CIAR Scholar and Alfred P. Sloan Foundation Fellow}

\begin{abstract}
We examine the origin of clustercentric gradients in the star
formation rates and colors of rich cluster galaxies within the context
of a simple model where clusters are built through the ongoing
accretion of field galaxies. The model assumes that after galaxies
enter the cluster their star formation rates decline on a timescale of
a few Gyrs, the typical gas consumption timescale of disk galaxies in
the field. Such behaviour might be expected if tides and ram pressure
strip off the gaseous envelopes that normally fuel star formation in
spirals over a Hubble time. Combining these timescales with mass
accretion histories derived from N-body simulations of cluster
formation in a $\Lambda$CDM universe, we reproduce the systematic
differences observed in the color distribution of cluster and field
galaxies, as well as the strong suppression of star formation in
cluster galaxies and its dependence on clustercentric radius.  The
simulations also indicate that a significant fraction of galaxies
beyond the virial radius of the cluster may have been within the main
body of the cluster in the past, a result that explains naturally why
star formation in the outskirts of clusters (and as far out as two
virial radii) is systematically suppressed relative to the field.  The
agreement with the data beyond the cluster virial radius is also
improved if we assume that stripping happens within lower mass
systems, before the galaxy is accreted into the main body of the
cluster.  We conclude that the star formation rates of cluster
galaxies depend primarily on the time elapsed since their accretion
onto massive virialized systems, and that the cessation of star
formation may have taken place gradually over a few Gyrs.
\end{abstract}

\section{Introduction}
Extensive observational work has established that the star formation
properties of galaxies in rich clusters differ significantly from
those of field galaxies (e.g., \cite{O60,DTS,B+97,KK}). In the field,
galaxies form stars at rates several times higher than systems of
similar luminosity in the cores of clusters. This is partly a result
of the well-known morphology-density relation, since ellipticals and
S0 galaxies are more abundant in clusters (\cite{Dressler,WGJ}), but
there is evidence that even later type galaxies in clusters form stars
at lower rates then in the field.  For example, Balogh et al
(\cite*{PSG}) report that field galaxies of given bulge-to-disk ratio
and luminosity have, on average, much larger [OII] equivalent widths
than their counterparts in rich clusters, suggesting that the cluster
environment somehow curbs the star formation rates of all galaxies,
regardless of morphology.

Various physical mechanisms have been proposed to explain this and
other systematic differences between the field and cluster galaxy
populations. Ram pressure stripping by the intracluster medium, for
example, has been suggested as a means of removing the gaseous
component of disk galaxies and of dramatically altering their
morphology and subsequent star forming history (e.g.,
\cite{GG,FN99,A+99}). Tides, either by the main cluster potential
(e.g. \cite{BV,F98}) or by fly-by encounters with other cluster
galaxies (``galaxy harassment'', \cite{Moore+96}), have also been
proposed to explain the observations.

Although the above processes are all plausible, the actual changes in
star formation rate induced by them remain a matter of debate. For
example, some authors have suggested that the cluster environment
triggers intense bursts of star formation that rapidly consume the gas
of an infalling cluster galaxy (\cite{DG83,Barger,P+99}), while others
have favored a scenario where star formation is truncated in a galaxy
almost immediately after it is accreted into the cluster
(\cite{A2390,NBK,1621,JSC}).

The virtues of these scenarios can in principle be tested
observationally, since they are expected to produce a population of
galaxies with unusually strong Balmer absorption lines in which star
formation has been recently terminated (e.g., \cite{DG83,CS87}).
Recently, Balogh et al. (\cite*{PSG}) have applied this idea to
galaxies in the CNOC1 survey of X-ray luminous clusters
(\cite{YEC}). From the paucity of galaxies with strong H$\delta$
absorption in their spectra they conclude that the decline of star
formation in these clusters may actually be a fairly gradual process.
This is in contradiction to other authors' conclusions, which are
derived from datasets with relatively high numbers of H$\delta$-strong
objects (e.g. \cite{Barger,P+99}); whether this discrepancy is due to
selection effects or to real differences in the clusters studied has
not yet been fully elucidated.

A natural timescale for a gentler reduction in star formation rate may
be gleaned from the long--known observation that, given their present
disk gas content, normal field spirals currently form stars at rates
that cannot be sustained over a Hubble time. At face value, it appears
that most spirals would use up their disk gas supply in a few Gyrs
(e.g., \cite{G+89}), though this timescale may be considerably
extended for some choices of the IMF and if the effects of
non-instantaneous mass recycling is taken into account (\cite{K+94}).
Star formation lifetimes can be significantly longer if galaxies
continuously accrete fresh star formation fuel from their surroundings
(e.g., \cite{HVC}). If, as proposed by Larson, Tinsley \& Caldwell
(\cite*{LTC}), this extended reservoir is stripped off a galaxy when
it first enters the cluster, its star formation rate may decay
significantly within a few Gyr, leading to large differences in the
cluster and field populations. In clusters that are built
hierarchically (as in the ``bottom-up'' scenario favored by cold dark
matter cosmogonies) this mechanism would also establish a radial
gradient in the star formation properties of cluster galaxies,
reflecting the relation between the clustercentric radius of a galaxy
and the time of its accretion into the cluster. This is an important
ingredient in the success of semianalytic models, which have been
shown to match global cluster properties such as the morphological
composition and the blue galaxy fraction (\cite{semianal,KC}).

We explore here a simple model based on this interpretation where the mass
accretion history of a cluster obtained from numerical simulations of a universe
dominated by cold dark matter is coupled with a simple star formation
prescription for galaxies following accretion. Once calibrated to reproduce the
properties of local field galaxies, the model has no free parameters and its
results can be compared directly with observations. We shall focus our analysis
on a quantitative discussion of the clustercentric radial gradients in star
formation properties and galaxy colors expected in this model.

In \S\ref{sec-sample} we describe the observational dataset we use, selected
from the CNOC1 survey. The numerical simulations used to derive mass accretion
histories of different clusters are described in \S\ref{sec-sims}.  Our model
prescriptions are presented in \S\ref{sec-model} and our results in
\S\ref{sec-results}.  We discuss the implications of our results in
\S\ref{sec-discuss}, and list our conclusions in \S\ref{sec-conc}.

\section{Observational Dataset}\label{sec-sample}
We use in this paper the CNOC1 cluster redshift survey dataset (\cite{YEC}),
which consists of CFHT spectra for $\sim 2000$ galaxies in 15 X--ray luminous
clusters at $0.19<z<0.55$.  The observational selection effects of this survey
are well understood and are discussed in detail in Yee, Ellingson \& Carlberg
(\cite*{YEC}) and Balogh et al. (\cite*{PSG}).  For the present analysis, we
weigh the raw data by three factors: one to account for the primary selection
effect due to source magnitude, and two secondary corrections which depend on
the galaxy color and position on the CCD.  One of the main advantages of this
survey, especially for the purposes of the present study, is that the dataset
includes spectra of foreground and background field galaxies projected onto each
cluster.  Since both field and cluster galaxies are selected using the same
criteria, {\it relative} differences between the two samples are rather
insensitive to uncertainties in the procedure used to correct for selection
effects.

Since our analysis concentrates on radial gradients within clusters,
we only consider those clusters with well defined centers in position
and velocity space and thus exclude from the sample the bimodal
clusters MS0906+11 and MS1358+62 (\cite{CNOC1}).  We also restrict the
redshift range of the galaxy sample to $0.19<z<0.45$, in order to
facilitate comparisons with simulations analyzed at a single epoch and
to minimize effects due to global changes in the galaxy population as
a function of redshift. This effectively removes two more clusters
from the CNOC1 sample, MS0016+16 and MS0451-03.  The final sample has
twelve clusters in total, which are scaled and co-added together to
construct a ``fiducial cluster'' sample where effects due to
substructure and asphericity of individual clusters are minimized
(\cite{CYE}). The full procedure is described in detail in Balogh et
al (\cite*{B+97,PSG}). 

In brief, we use the mass models of Carlberg, Yee \& Ellingson
(\cite*{CYE}) to divide the sample into a cluster and field sample;
galaxies are deemed to be cluster members if they are within 3$\sigma$
of the (radially dependent) cluster velocity dispersion, and field
members if they are beyond 6$\sigma$.  Cluster galaxy positions are
all measured relative to the brightest cluster galaxy (BCG), and
normalized to $R_{200}$, the radius at which the mean inner density is
200 times the critical density. The BCGs themselves are omitted from
the final sample, as they are likely to have a unique formation
history which may differ from the general cluster population; we
briefly compare their properties with those of the full sample in
\S\ref{sec-nevsfr}. Finally, we impose an absolute magnitude limit on
the sample, considering only galaxies brighter than
$M_r=-18.5+5\log{h}$ at $z=0$ (Gunn-$r$, $q_0=0.1$); when
appropriately weighted, this sample is statistically complete. Because
each individual cluster is at a different redshift a small
evolutionary correction is applied to this cutoff assuming that
luminosity increases in direct proportion to $(1+z)$ (e.g.,
\cite{CNOC2}).  At $z \sim 0.3$, the luminosity cutoff we adopt is
therefore $M_r=-18.8+5\log{h}$.  This correction has little effect on
our results because of the narrow redshift range under consideration.
For the sample considered here, the luminosity function of the cluster
population is similar to that of the field galaxy population.

Individual star formation rates (SFRs) for galaxies in the sample are computed
from the rest frame equivalent width of the \oii\ emission line and the rest
frame B--band luminosity relative to solar ($L_B/L_{B,\odot}$) using the relation,
\begin{equation}\label{eqn-sfr}
\dot{M_*}=3.4 \times 10^{-12} \left({L_B \over L_{B,\odot}}\right)
W_\circ(\rm{OII}) \, E(H\alpha) \, M_\odot \, \mbox \, {  yr}^{-1}.
\end{equation}
Here E(H$\alpha$) is the extinction at H$\alpha$ which, following Kennicutt
(\cite*{K92}), we take to be one magnitude.  The coefficient in Equation
\ref{eqn-sfr} has been chosen so that, on average, the SFRs of field
galaxies are consistent with their colors and luminosities, based on models of
their star formation history that are discussed in detail in \S\ref{sec-evsfr}.
This coefficient is consistent with that empirically determined by Barbaro \&
Poggianti (\cite*{BP}), and about 30\% larger than that measured by Kennicutt
(\cite*{K92}) --- well within the uncertainty of its determination (see
Kennicutt \cite*{Kenn_review}).  Because we are concerned with relative
differences, neither this normalization nor the extinction correction has a
significant effect on our conclusions, unless these quantities vary dramatically
from the cluster to the field.

Uncertainties in the equivalent widths have been assessed assuming Poisson
statistics, and have been internally calibrated to account for additional
systematic errors, as described in Balogh et al. (\cite*{PSG}).  We exclude from
the sample all galaxies with very large \ow \ uncertainties, $\Delta$ \ow $>
15$\AA\ ($\sim 6 \%$ of the sample), and all galaxies for which \oii\ lies outside
the observed spectral range.  The final sample with measured SFRs consists of
556 cluster galaxies and 339 field galaxies.

The \ow\ measurements from which we derive SFRs are computed by adding up the
observed flux (accounting for partial pixels) above the continuum level in the
wavelength range $3713<\lambda/$\AA$<3741$. The continuum level is estimated by
fitting a straight line to the flux in the ranges $3653<\lambda/$\AA$<3713$ and
$3741<\lambda/$\AA$<3801$. For weak or absent [OII] features, the \ow\ index
will be sensitive to features in the continuum in these two regions. A crude
estimate of uncertainties introduced in the \ow \ measurements by these features
may be obtained by using eq.~1 to compute SFRs of galaxies expected to have
little or no ongoing star formation; these are red cluster galaxies with large
4000\AA\ breaks ($(g-r)_\circ>0.35$, $D_{4000}>1.8$) found within
$0<R/R_{200}<0.3$. The 3$\sigma$--clipped mean SFR of this population is -0.057
$h^{-2} \, M_\odot \, \mbox{yr}^{-1}$, and the standard deviation is $0.156 \,
h^{-2} \, M_\odot \, \mbox{yr}^{-1}$. This small systematic offset and
uncertainty are taken into account in our modeling, as described in
\S\ref{sec-sfmodel}.

\section{Numerical Simulations}\label{sec-sims}

Cluster mass accretion rates are computed directly from N--body simulations of
the formation of six massive clusters ($0.7<M/(10^{15}M_\odot)<2.3$) in a
COBE-normalized, $\Lambda=0.7$, $\Omega_0=0.3$ cosmology.  The simulations are
similar to those described in detail by Eke, Navarro, \& Frenk (\cite*{ENF}).
We assume that ``light traces mass'' in the clusters and identify statistically
each dark matter particle with a ``galaxy''. The virial radius, $R_{\rm vir}$,
of a cluster is computed using the overdensity prescription described in Eke,
Cole \& Frenk (\cite*{ECF}), which differs slightly from $R_{200}$.  At $z=0.3$,
the mean redshift of our cluster sample, $R_{\rm vir} \approx 1.2 R_{200}
\approx 1.4$-$2.4$ Mpc. Each cluster has about $9,000$ particles within $2\,
R_{\rm vir}$ at $z=0.3$. The observations and model parameters are all presented
in terms of the simulation cosmology; we use $H_0=70$ km s$^{-1}$ Mpc$^{-1}$,
which gives a present age of the universe of 13.5 Gyr. At $z=0.3$ the universe
is $\sim 11$ Gyr old.

\section{The Model}\label{sec-model}
As discussed in \S1, our modeling assumes that cluster galaxies differ from
their field counterparts in their star forming properties because they are
stripped of their surrounding gas reservoirs as they are accreted into the
cluster. Star formation in cluster galaxies thus declines after accretion as
more and more of the galaxy's remaining gas content gets turned into stars. 

The model involves the following steps: 

\items{
(i) All particles within $2 \, R_{200}$ from the center of each simulated
cluster at $z\sim 0.3$ are selected.}

\items{(ii) Each particle is traced back in time to find out when it was first accreted
into the cluster. Two definitions of ``accretion time'', $t_{\rm acc}$, are
used.  The first ($t_{\rm acc}=t_{\rm cluster}$) is defined to be the time when
the particle (``galaxy'') first finds itself within $R_{\rm vir}$ from the
center of the current most massive progenitor of the cluster, and the second
($t_{\rm acc}=t_{\rm group}$) is the time when a galaxy is first accreted into
any large clump, not necessarily the most massive one. Details of this procedure
are given in \S\ref{sec-tacc}.}

\items{
(iii) A SFR is chosen for each galaxy at $t=t_{\rm acc}$. For simplicity, these
are taken at random (using appropriate weights as discussed in
\S\ref{sec-sample}) from the distribution of SFRs computed for $z\sim 0.3$ field
galaxies in our sample. This assumes implicitly that field SFRs do not evolve
with time, arguably the simplest possible model. We explore in \S\ref{sec-evsfr}
the consequences of relaxing this assumption.}

\items{
(iv) We use a simple gas consumption model to estimate SFRs following accretion
and to compute the SFR of each cluster galaxy at $z \sim 0.3$. Details are
discussed in \S\ref{sec-sfmodel}.}

\items{
(v) Each simulated cluster is projected onto three orthogonal planes.  Field
contamination in the observational sample is accounted for as discussed in
\S\ref{sec-fieldcon}.}

This procedure uniquely defines a final ($z=0.3$) SFR for each galaxy in the
simulated clusters; the mean SFR can then be computed as a function of
clustercentric distance, and compared with the field galaxy observations.  We
discuss now the various steps of the model in some detail.

\subsection{Accretion Times}\label{sec-tacc}
We have explored two ways of assigning ``accretion times'', $t_{\rm
acc}$, to galaxies in the simulated clusters.  One choice defines
$t_{\rm acc}=t_{\rm cluster}$ as the time when a particle is first
found within the virial radius of the most massive progenitor present
at that time. Following Eke, Cole \& Frenk (\cite*{ECF}), the virial
radius, $R_{\rm vir}$, is defined as the radius where the mean inner
density of the cluster is $\bar{\rho}(R_{\rm
vir})=\Delta_c(z)\rho_c(z)$, where $\rho_c(z)$ is the critical density
at the redshift $z$, and $\Delta_c(\Omega(z))$ is the ``critical''
overdensity for spherical collapse (which takes the familiar value of
178 for $\Omega=1$ models).

The above definition of $t_{\rm cluster}$ may not be completely appropriate, since
in a hierarchically clustering universe particles may be first accreted into
another ``protocluster'' before being accreted into the main progenitor of the
final cluster. Therefore, we consider as an alternative definition of $t_{\rm
acc}$ the time when a particle first finds itself associated with a clump with
circular velocity exceeding $V_c=500$ km s$^{-1}$. (This circular velocity
corresponds to a virial temperature of $\sim 0.8$ keV). Particles are associated
with clumps via a friends-of-friends algorithm, with an evolving linking length
parameter taken to be $10\%$ of the mean inter-particle separation at each time.
Accretion times defined this way are labeled $t_{\rm group}$. We shall see in
\S\ref{sec-nevsfr} that our results are rather insensitive to the
particular choice of accretion time definition.

The ``resolution'' of $t_{\rm acc}$ is limited by the number of times particle
positions are output by the simulations.  In this case, we have outputs every
1.34 Gyr; we therefore add a uniform random number between 0 and 1.34 Gyr to each
particle's $t_{\rm acc}$ to smooth out this resolution.  We do not expect our
results to be sensitive to this resolution, since we are combining the results
of six clusters, which will smooth out the effect of discrete merger events.

\subsection{Star Formation Prescription}\label{sec-sfmodel}

For normal, field spiral galaxies, the SFR per unit area averaged over the disk,
$\Sigma_{\rm SFR}$, depends on the disk gas content in a manner well
approximated by a Schmidt law (\cite{Schmidt}).  From Kennicutt (\cite*{K98}),
this relation is given by
\begin{equation}
\Sigma_{\rm SFR}=(2.5\pm0.7)\times10^{-4}\left(\Sigma_{\rm gas} \over  M_\odot
\mbox{pc}^{-2}\right)^N M_\odot \, \mbox{yr}^{-1} \, \mbox{kpc}^{-2},
%\dot{M_*}(t)=K M_g^N=-{dM_g \over dt},
\end{equation}
where $\Sigma_{\rm gas}$ is the average gas surface density, and the
exponent $N=1.4 \pm 0.15$.  Converting surface densities to integrated
values using the source diameters tabulated by Kennicutt, galaxies that evolve
according to this relation and accrete no extra gas would form stars
at a steadily decreasing rate given by,
\begin{equation}\label{eqn-sfrt}
\dot{M_*}(t^\prime)=
\dot{M_*}(0)\left(1+{0.33 {t^\prime \over t_e}}\right)^{-3.5} M_\odot \mbox{yr}^{-1},
\end{equation}

where $\dot{M_*}(0)$ is the initial SFR and $t_e \approx 1.48 \,
(\dot{M_*}(0)/M_{\odot}\, {\rm yr}^{-1})^{-0.29}$ Gyr is the
characteristic gas consumption timescale.  As discussed in
\S\ref{sec-model}, we adopt $\dot{M_*}(0)$ values taken at random from
the measured field SFRs and set $t^\prime=t-t_{\rm acc}$. With these
assumptions it is possible to compute SFRs of cluster galaxies at
$z\sim 0.3$, the mean redshift of the observational sample.  In the
above calculation, we have neglected the effects of gas recycling.  If
we assume instantaneous recycling, where a fraction $R$ of every solar
mass of stars formed is returned as gas, the timescale $t_e$ is
increased by a factor $1/(1-R)$.  Assuming $R=0.33$
(\cite{K+94})\footnote{This large factor of $R$ approximates the
effects of a non-instantaneous recycling calculation, assuming a Scalo IMF,
and is fairly good for slowly evolving disks; for rapidly evolving disks, $R$ can be
as high as 0.75.}, we obtain $t_e \approx 2.2 \,
(\dot{M_*}(0)/M_{\odot}\, {\rm yr}^{-1})^{-0.29}$ Gyr. This increase
does not have a significant effect on the results we present, so we
neglect the effects of recycling throughout this work.

Finally, we correct the values obtained from eq.~\ref{eqn-sfrt} for the
systematic offset and uncertainty in the SFR measurements discussed in
\S\ref{sec-sample}.  This is done by adding a  random number, drawn from
a Gaussian distribution with a mean of $-0.12 M_{\odot}$ yr$^{-1}$ and a
variance of $0.32 M_{\odot}$ yr$^{-1}$, to the model SFR.  These parameters were
chosen from the SFR distribution of red, central cluster galaxies (with
$h=0.7$), as described in \S\ref{sec-sample}.

\subsection{Field Contamination}\label{sec-fieldcon}

Observational cluster datasets are contaminated by field galaxies, projected
onto the cluster, that have Hubble-flow redshifts similar to the peculiar
velocities induced by the cluster potential. Our model takes this into account
following the procedure of Carlberg et al. (\cite*{CYE}). In brief, the velocity
differences between cluster and individual galaxies in each dataset are scaled
to the velocity dispersion of each cluster and their clustercentric projected
radii to $R_{200}$. These normalized velocities, $V_{\rm norm}=\Delta v/\sigma$,
and radii, $R_{\rm proj}/R_{200}$, are then combined into a single dataset.  The
density of galaxies per unit $V_{\rm norm}$ at $5<V_{\rm norm}<25$ is computed
in radial bins to account for the radially varying $\sigma$.  Under the
assumption that this density remains constant within each bin, it is used to
calculate the expected number of interloper galaxies, i.e., those within
3$\sigma$ in each radial bin.  This provides a direct estimate of the fraction
of galaxies deemed cluster members that are actually field galaxies projected
onto the cluster in ``redshift space''. This fraction is typically $\sim 1\%$ in
the central cluster regions, but may be as large as $\sim 12\%$ at $R_{200}$.
Our modeling accounts for this effect by including an appropriate number of
``field'' galaxies (randomly selected from the field sample) in all
computations.

\section{Results}\label{sec-results}
\subsection{Non-evolving field SFR model}\label{sec-nevsfr}

\begin{figure*}
\begin{center}
\leavevmode \epsfysize=8cm \epsfbox{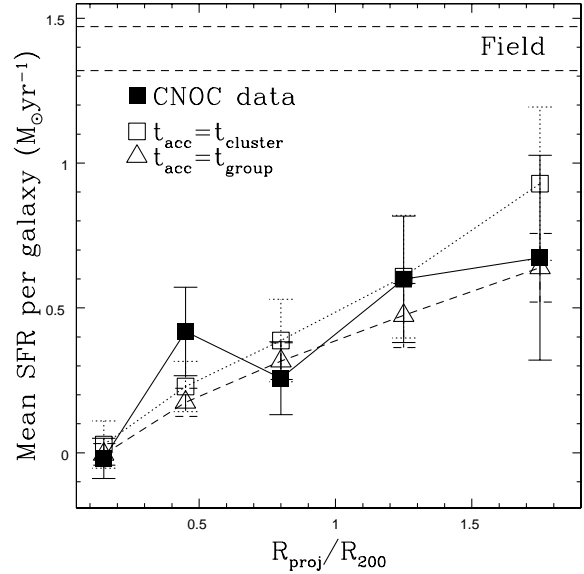}
\end{center}
\caption{
The mean SFR per galaxy as a function of projected radius for galaxies in the
CNOC1 cluster sample (solid squares) compared with the model predictions under
the assumption that $t_{\rm acc}=t_{\rm cluster}$ (open squares) and that
$t_{\rm acc}=t_{\rm group}$ (open triangles).  The horizontal dashed lines
correspond to the field SFR, bracketed by its 1-$\sigma$ dispersion. The SFR
gradient in the CNOC1 clusters is accurately reproduced by our simple accretion
models.  Error bars are all 1$\sigma$.
\label{fig-fig1}}
\end{figure*}

Figure \ref{fig-fig1} shows the mean SFR per galaxy as a function of projected
distance from the cluster center, normalized to $R_{200}$. The solid squares
correspond to the CNOC1 data for the complete sample of galaxies brighter than
$M_r=-19.6$ (at $z=0.3, h=0.7$), averaged over radial bins and with 1-$\sigma$
jackknife error bars. The mean SFR per galaxy increases systematically from the
center of the cluster outwards, from almost zero near the center to about $\sim
0.7 \, M_{\odot}$ yr$^{-1}$ in the outskirts of the cluster. We note that the
absolute SFR values quoted here are sensitive to a number of sample selection
parameters; in particular to the luminosity cutoff and the uncertain coefficient
in eq.~\ref{eqn-sfr}, so care must be exercised when comparing these results to
other work. On the other hand, relative differences between cluster and field
populations are robust, since the two samples have similar luminosity functions,
and are drawn from the same survey with identical selection criteria.

As shown in Figure \ref{fig-fig1}, the average SFR per galaxy in the
field is significantly higher than in clusters. Remarkably, even at
radii as far from the center of the cluster as $\sim 2 R_{200}$,
cluster star formation rates remain depressed by almost a factor of
two relative to the field.\footnote{Because the luminosity functions
of the field and cluster population considered in this sample are
quite similar, the same result is obtained for the SFR per unit
luminosity as for the SFR per galaxy.}.  We note here that the
BCGs, which are omitted from the data sample, have unusually high
SFRs; the mean SFR of the 11 BCGs which satisfy our selection
criteria\footnote{No spectrum is available for the BCG in MS
0451.5+0250.} is 14 $M_\odot$ yr$^{-1}$, much greater than the mean
field value.  Only four of these eleven galaxies have no significant
[OII] emission.  The low redshift BCGs are discussed in more detail in
Davidge \& Grinder (\cite*{DG95}); we do not consider these galaxies
further in this study.

Figure \ref{fig-fig1} also shows, with open symbols, the results of the modeling
procedure outlined in \S\ref{sec-model}. Open squares and triangles correspond
to the two different accretion time definitions discussed in
\S\ref{sec-tacc}. Error bars represent the 1-$\sigma$ variance of the 18
numerical realizations (6 simulated clusters and 3 orthogonal projections per
cluster). The agreement between the model and the observations is remarkable,
especially considering that the modeling involves no free parameters. The
model reproduces the observed SFR gradient and even the observed depression,
relative to the field, of SFRs outside $R_{200}$.

The latter result is somewhat surprising, since in spherical accretion models
particles outside $R_{200}$ are infalling into the cluster for the first time
(\cite{B85,WNEF93}) and therefore their SFRs have yet to feel the
effects of the cluster environment. The main reason for our result is that a
substantial fraction ($54 \pm 20 \%$ for the six clusters we studied) of
particles between $1$ and $2 \, R_{200}$ have actually been inside the virial
radius of the main progenitor at some earlier time.  These are often particles
that populated the outskirts of recently accreted clumps and that, although
still bound to the system, have been scattered to large apocenter orbits during
the merger process.

Interestingly, assuming that the onset of the SFR decline occurs when a galaxy
is accreted into {\it any} clump with circular velocity exceeding $500$ km
s$^{-1}$ rather than the cluster's main progenitor has only a small effect on
this result within $R_{200}$ (witness the good agreement between open triangles
and squares in Figure \ref{fig-fig1}). Note that the mean SFR beyond $R_{200}$
is further suppressed under this assumption, resulting in even better agreement
with observations.  This is because some particles beyond $R_{200}$ are found
within fairly massive groups, although they may never have been within the
virial radius of the main cluster.

An important feature of this model is that many cluster galaxies at $z=0.3$ have
substantial SFRs.  In particular, near $R_{200}$, $\sim 20 \%$ of cluster
galaxies have SFRs in excess of 1$M_\odot \mbox{yr}^{-1}$, and this declines to
about $10\%$ at $R = 0.5 R_{200}$; fractions which are consistent with the CNOC1
data.  These large SFRs are not the result of cluster-induced starbursts, but
correspond to recently accreted field galaxies in which star formation has not
yet been completely quenched.  In a recent study, Balogh \& Morris
(\cite*{Halpha}) have measured H$\alpha$ equivalent widths for galaxies in Abell
2390, and they failed to find a substantial population with large H$\alpha$
fluxes that were undetected in [OII]. Thus, there does not appear to be a
population of dust-obscured starburst galaxies in this cluster which were missed
in the CNOC1 survey.  Our modeling indicates that cluster-induced starbursts are
not necessary to generate the levels of star formation seen in these clusters.

We conclude from this comparison that a model based on the simple assumption
that the SFR of a galaxy decreases on gas consumption timescales is able to
reproduce the data extremely well, lending support to the underlying hypothesis
that continuous accretion of external gas is responsible for maintaining the
SFRs of normal spirals over a Hubble time.

\subsection{Evolving field SFR model}\label{sec-evsfr}

There are two weaknesses in the model explored in the previous sections. One is
the assumption that the SFRs of field galaxies, which are used to assign
``initial'' SFRs to cluster galaxies at accretion time (see eq.~\ref{eqn-sfrt}),
do not evolve with time.  Given the mounting evidence that the average SFR per
unit volume evolves strongly with lookback time (\cite{CFRS6,Madau}) it is
necessary to explore the consequences of relaxing the non-evolving SFR
assumption on our results. The second is the luminosity evolution of cluster
galaxies, which must also be modeled in order to account for galaxies that may
fade beyond the observational magnitude limit when their SFR declines.

\begin{figure*}
\begin{center}
\leavevmode \epsfysize=8cm \epsfbox{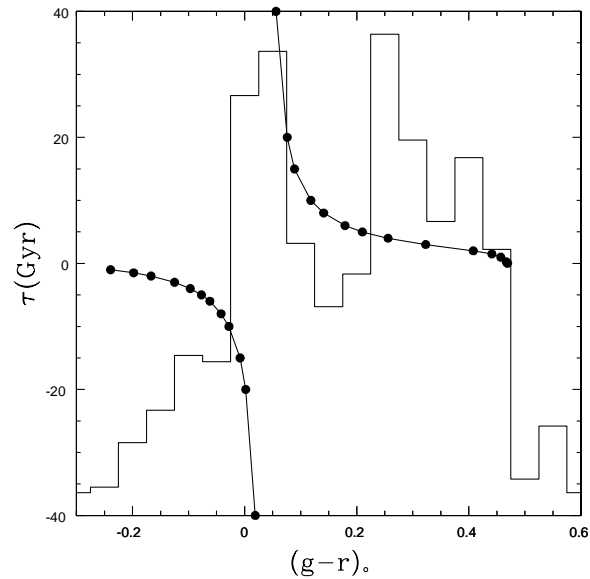}
\end{center}
\caption{
The histogram shows the distribution of $(g-r)_\circ$ colors for the field
galaxy sample, in arbitrary units. The solid symbols correspond to model colors
at $z=0.3$ (an age of 11 Gyr) for galaxies with star formation history given by
eqn. \ref{eqn-tau} and parameter $\tau$ as shown on the ordinate axis.  The
solid line through these points is the linear interpolation used to assign
$\tau$ to each galaxy, based on its observed $(g-r)_\circ$.
\label{fig-fig1_2}}
\end{figure*}

A proper treatment of these effects, which must take into account the merger and
star formation history of galaxies, is of great interest, but well beyond the
scope of this study.  In order to at least explore the effects that this more
complete treatment will have, we use simple $\tau-$models to model the SFR
evolution of field galaxies in our sample, i.e., by assuming that,
\begin{equation}\label{eqn-tau}
\dot{M_*}(t)=\dot{M_*}(t_0) \, {\rm e}^{-(t-t_0)/\tau},
\end{equation}
where $\dot{M_*}(t_0)$ corresponds to the SFR of galaxies in the field at
$t=t_0\approx 11$ Gyr, the age of the universe at $z\sim 0.3$. Once $\tau$ is
determined for each galaxy it is possible to construct, at arbitrary times, an
``evolving'' field SFR distribution which matches the observations at
$z\sim 0.3$. For a given IMF, and constant reddening, $\tau$ is uniquely determined
by the observed $(g-r)_\circ$ color 
\footnote{
This color is available for all galaxies in the CNOC1 dataset. We have made a
small adjustment to the zero point of the observed $(g-r)_\circ$ colors,
reducing them by $0.05$ so that the reddest observed galaxies have the colors
predicted by a model in which all stars formed at $t=0$ in an instantaneous
burst.}
of the galaxy, as shown in Figure \ref{fig-fig1_2}.  We have used the
PEGASE (\cite{PEGASE}) spectrophotometric models with a Salpeter IMF
(with lower and upper limit given by $0.1<M/M_\odot<120$) to determine
$\tau$ for each galaxy in our field sample.  The bluest galaxies
($(g-r)_\circ<0.04$) have $\tau<0$, corresponding to a SFR that
increases with time, while moderately blue galaxies ($(g-r)_\circ
\approx 0$) have $\tau$ approaching (positive or negative) infinity,
corresponding to a constant SFR.  The reddest field galaxies, those
with $\tau<300$ Myr, are assumed to have formed in a single burst at
high redshift, and are modeled as such in the cluster.  The net result
is a rather extreme model where the global star formation rate in the
field population increases steadily out to $z\approx 10$. We explore
below as well the consequences of restricting this increase at large
$z$, and conclude that our results are quite insensitive to the
precise nature of the redshift evolution in the field.

\begin{figure*}
\begin{center}
\leavevmode \epsfysize=12cm \epsfbox{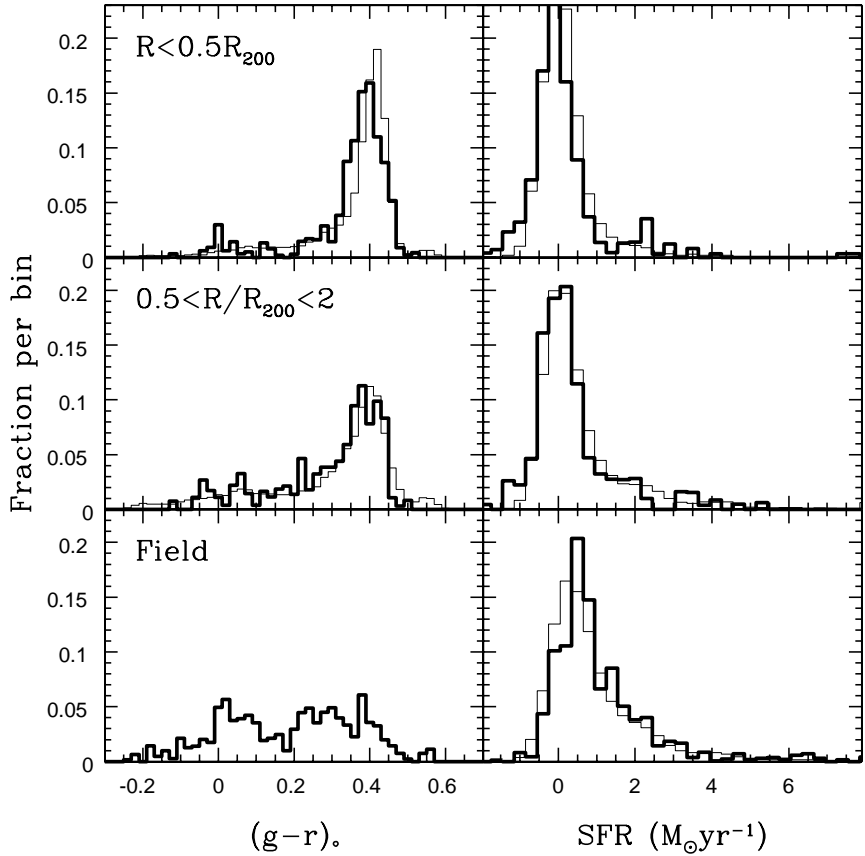}
\end{center}
\caption{
{\it Left panels: }$(g-r)_\circ$ distributions in the observed CNOC1 sample
({\it heavy solid lines}) and our evolving field-SFR model ({\it thin solid
lines}).  The samples are divided into a field sample and two cluster samples
(inner and outer regions), as labeled in each panel.  By construction, the model
color distribution matches the field observations exactly.  {\it Right panels: }
SFR distributions for the same samples shown in the left panels.  The model
provides a good match to the color and SFR distributions of the observed cluster
population, although it slightly overestimates the mean cluster SFR.
\label{fig-fig2}}
\end{figure*}

The normalization constant of eq.~\ref{eqn-tau}, $\dot{M_*}(t_0)$, may in
principle be determined by the {\it observed} field galaxy SFR at $t_0$, but we
found that, because of the large uncertainties in individual SFR determinations,
this procedure leads to large discrepancies in the total luminosity of galaxies
in the sample once the SFRs are integrated over time. Because total luminosities
are considerably better determined than SFRs in our dataset, and are less
sensitive to burst--like events that eq.~\ref{eqn-tau} cannot reproduce, we
compute $\dot{M_*}(t_0)$ for each field galaxy by requiring the model to
reproduce its observed total Gunn-$r$ luminosity at $z=0.3$.  This choice
implies that, on a galaxy by galaxy basis, the model $z=0.3$ SFRs no longer
correspond to those derived from the \ow\ measurements.  The observed and model
field SFR distributions are compared in the bottom right panel of Figure
\ref{fig-fig2}; where the coefficient in eq.~\ref{eqn-sfr} ($3.4 \times 10^{-12}$) 
has been chosen so that
both distributions have identical averages. 
Although the difference is statistically significant, the two distributions are
qualitatively fairly similar, indicating that, on average, the simple
$\tau$-model provides a reasonable description of the $z=0.3$ field.  This
consistency allows us to compare model and observed SFRs, even though the former
is based on galaxy color, while the latter is determined from nebular emission.

A further advantage of modeling the SFR history of the field population is that
it allows us to model the color evolution of cluster galaxies.  Color
($(g-r)_\circ$) distributions of model galaxies (which include the addition of a
Gaussian random number to account for the $0.02$ magnitude error typical of the CNOC1
data) are compared with the observational data in the left panels of Figure
\ref{fig-fig2}.  The model shown adopts $t_{\rm acc}=t_{\rm
group}$, and excludes galaxies fainter than $M_r=-19.6$ at $z=0.3$.  
This model is very successful at producing a cluster core
dominated by red galaxies from an initial field galaxy sample that has a much
broader and bluer color distribution.  In the outskirts of the cluster, both the
observed and model color distributions show a significant increase in the
population of blue galaxies.  The qualitative agreement between model and
observed SFR and color distributions throughout the cluster is quite remarkable
for such a simple prescription.

The right panels of Figure \ref{fig-fig2} compare the observed
SFR distributions with this same model. In agreement
with observations, model distributions show a strong reduction in the fraction
of galaxies forming stars at rates higher than about $\sim 1 M_{\odot}$
yr$^{-1}$ near the cluster center. The agreement between model and observations
is quite good, although model galaxies form stars at slightly higher rates in
both bins.

\begin{figure*}
\begin{center}
\leavevmode \epsfysize=8cm \epsfbox{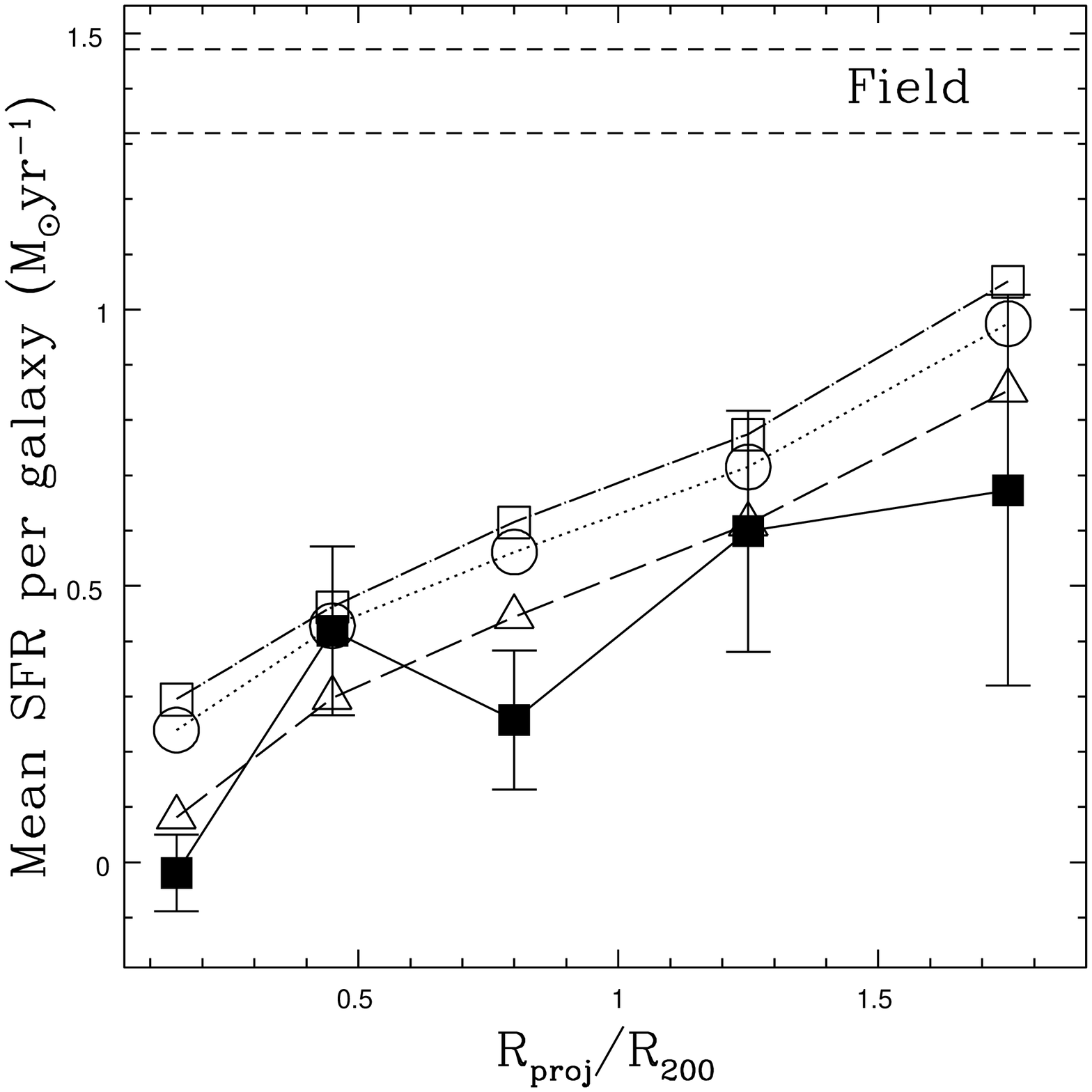}
\end{center}
\caption{
Same as Figure \ref{fig-fig1} but for three evolving field SFR models.  In the
fiducial model (open circles), the SFRs in field galaxies evolve as dictated by
simple $\tau$ models normalized to match the colors of field galaxies at $z\sim
0.3$, and galaxies that fade below the luminosity limit are excluded (see
\S\ref{sec-evsfr}). Accretion times are defined as $t_{\rm acc}=t_{\rm group}$.
The open squares show the result of modifying the evolution such that each
galaxy's SFR remains constant at $z>1.5$.  Neglecting the effects of galaxy
fading in the fiducial model results in the open triangles.  The data are shown
by the filled squares, connected by the solid line.
\label{fig-fig3}}
\end{figure*}

The differences between model and observed mean cluster SFRs are
clearly apparent in Figure \ref{fig-fig3}, which shows the
clustercentric gradient predicted by three variants of the evolving
field-SFR model.  The open circles correspond to the fiducial model as
described above, and are generally larger than the observed cluster
mean.  The open triangles show how this result changes if we neglect
galaxy fading; i.e., when we include all cluster galaxies in the
average.  This recovers the good match to the data, and shows that it
is primarily the luminosity fading of galaxies that affects our
result, and not the nature of the redshift evolution itself.  To
demonstrate this explicitly, we present a model (open squares) in
which the field galaxy SFRs are evolved as in the fiducial model but
are held constant for $z>1.5$.  The results obtained with this model
do not differ significantly from those of the fiducial model, lending
support to our conclusion that our results depend only mildly on the
SFR redshift evolution.

%The small mismatch of the evolving-SFR model to the data is thus due to the
%effect of luminosity fading.  In particular, galaxies which are accreted at
%early times, which would be among the reddest at the final epoch, have not
%formed many stars by their accretion time and are therefore too faint to make it
%into the final sample.  
%This problem is exacerbated in the case of the bluest galaxies; many of these
%are likely to be undergoing short bursts of star formation that involve a small
%percentage of the total stellar mass, a possibility our models cannot account
%for. In this case, colors are dominated by the brightest, youngest stars and the
%assigned $\tau$ ($<0$) may be uncharacteristically short compared with the
%timescale on which the galaxy truly formed the bulk of its stars. Our model
%assumes that these galaxies formed {\it all} of their stars very recently; if
%such a galaxy is accreted into the cluster at early times it would never become
%bright enough to be included in our sample.  In a more realistic model, the bulk
%of their stars would still be in place at early times, and their early accretion
%into the cluster would imply that they would be galaxies with old stellar
%populations by $z=0.3$.  Thus, our procedure may inappropriately remove some
%galaxies with low present SFRs from the final model cluster, artificially
%biasing the mean.  

We wish to stress that the purpose of the above exercise is to assess
the sensitivity of our modeling to various assumptions about the
redshift dependence of the SFR of the field galaxy population rather
than to build a realistic model of field galaxy evolution.  We
conclude that the luminosity evolution of the field galaxies has a
small, but non-negligible effect on our results. Although it appears
that qualitatively our conclusions are safe, it is clear that
definitive answers will have to wait for a more realistic modeling of
the field galaxy star formation evolution, such as the one implemented
in semianalytic models of galaxy formation (e.g. \cite{semianal},
Diaferio et al.  in preparation).
%Given the simplicity of the model, and the general good
%agreement with the data even in Figure \ref{fig-fig3}, we do not think that the
%results of the evolving-SFR model reveal important differences from the simpler
%model of \S\ref{sec-nevsfr}, though a more realistic modeling of the field
%galaxy star formation history would be of course desirable.

\section{Discussion}\label{sec-discuss}

We present models of the clustercentric dependence of star formation and colors
of galaxies in rich clusters, under the following assumptions: (i) the cluster
galaxy population is built by the ongoing accretion of field galaxies, (ii) SFRs
in field galaxies are sustained by regular accretion of gas from their
surroundings, and (iii) reservoirs of fresh star formation fuel are lost as
galaxies plunge into the cluster potential.  

Within this context, our model provides support for a gradual decline
(over a timescale of a few Gyrs) in the star formation rates of
cluster galaxies after accretion.  Actually, results similar to those
presented in the previous section may be obtained if the SFR in {\it
all} cluster galaxies is assumed to decay exponentially after
accretion with fixed timescales $1\lesssim t_e\lesssim3$ Gyrs.
Decline timescales longer than $\sim 3$ Gyr lead to unacceptably large
star formation rates in model clusters at $z\sim 0.3$. On the other
hand, sharp truncation of star formation ($t_e \lesssim 1$ Gyr) would
result in too little star formation within clusters and, furthermore,
would lead to an abundant population of galaxies with strong Balmer
lines but no nebular emission lines (K+A galaxies), and these appear
to be rare in the very luminous X-ray clusters we study here
(\cite{PSG}). Larger fractions of K+A galaxies have been reported in
other cluster datasets, in particular by the MORPHS collaboration
(\cite{D+99,P+99}), but it is still unclear whether this apparent
disagreement is a result of the procedure used to select the
spectroscopic sample, the effects of dust obscuration, or perhaps a
genuine effect of the dependence of SFRs on other cluster properties
such as X-ray luminosity or temperature (\cite{PSG}). Further analysis
is required in order to assess whether the simple model we propose
here is applicable to clusters other than the relatively massive,
X-ray luminous systems targeted by the CNOC1 survey.

With this caveat, the success of our model strongly suggests that (i) gradients
in galaxy properties arise from gradients in accretion times and (ii) that the
cessation of star formation need not take place abruptly to explain the observed
SFRs of cluster galaxies. On the other hand, our understanding of the star
formation history of cluster galaxies is bound to remain incomplete until the
physical mechanism responsible for the decline of star formation in cluster
galaxies is fully elucidated.  The loss of external gas reservoirs advocated
here is attractive, but conclusive observational evidence that such reservoirs
exist in isolated field galaxies has been slow to emerge (\cite{B+99}, but see
\cite{HVC}). Other processes that may reduce star formation rates, such as
tides, harassment, and ram pressure tripping, operate on similar timescales once
a galaxy has been accreted into the cluster and it is therefore unlikely that
analysis of the kind we present here will be able to distinguish clearly amongst
them.  

Another major question that our models do not address is the origin of
gradients in galaxy morphology that parallel the color and star
formation gradients we focus on (e.g., \cite{Dressler,D+97,KK}). The
over-representation of ellipticals near the center of X-ray luminous
clusters may reflect higher merger rates between nearly equal mass
systems in systems that collapse early to form cluster cores, but this
is an issue that remains unexplored in our model. The construction of
large spectroscopical datasets that probe the dependence of galaxy
SFRs, colors and morphologies on cluster properties such as
concentration, richness, velocity dispersion and X-ray properties,
coupled with numerical and/or semianalytical models that treat
self-consistently the accretion history and dynamics (and
hydrodynamics) of galaxy and cluster formation, are probably the most
promising way to make substantial progress in the subject.

\section{Conclusions}\label{sec-conc}

We present a simple model to account for the systematic differences in the star
formation properties of galaxies in CNOC1 clusters and the field. The model
assumes that the cluster is built through the accretion of field galaxies whose
star formation rates gradually decline after entering the cluster as a result of
the removal, through tides or ram pressure, of the gaseous envelopes needed to
supply normal spirals with the fuel needed to sustain their present star
formation rates over a Hubble time. Once calibrated to reproduce observations of
nearby spirals the model has no free parameters. 

Using cluster mass accretion rates determined from N--body simulations
of cluster formation in a $\Lambda$CDM universe, our model is able to
reproduce qualitative and quantitative differences in the mean star
formation rates and colors between clusters and the field. The model
demonstrates that the origin of radial gradients in these properties
is the natural consequence of the strong correlation between radius
and accretion times which results from the hierarchical assembly of
the cluster.  Interestingly, the model also explains why star
formation in the outskirts of clusters is found to be almost a factor
of two below the field average as far out as twice the virial radius
of the cluster. This is a result of cluster members being pushed onto
highly eccentric, loosely bound orbits during major merger events.
Our results are robust to evolution in the star formation properties
of field galaxies, but somewhat more sensitive to the manner in which
fading of cluster galaxies is modeled.  Realistic modeling of the
field population star formation histories is required to fully
understand the consequences of this fading.

We conclude that the stripping of extended gaseous reservoirs by the cluster
environment and the gradual decline that follows gas consumption is likely to be
the main mechanism that differentiates the star formation properties of cluster
and field galaxies.

\acknowledgments 
We have made extensive use of the CNOC1 dataset of intermediate--redshift
clusters.  We are grateful to all the consortium members and to the CFHT staff
for their contributions to this project.  We thank the referee for a careful
reading of the manuscript, and for useful suggestions which improved this
paper.
In Victoria, MLB was supported by a
Natural Sciences and Engineering Research Council of Canada (NSERC) research
grant to C. J. Pritchet and by an NSERC postgraduate scholarship.  In Durham,
MLB is supported by a PPARC rolling grant for extragalactic astronomy and
cosmology.

\newpage
\bibliographystyle{astron_mlb}
\bibliography{ms_pp}

\end{document}